\documentclass{article}
\usepackage[top=30truemm,bottom=60truemm,left=30truemm,right=30truemm]{geometry}
\normalsize

\usepackage{color}%
\usepackage[normalem]{ulem}%
\usepackage{slashed}%
\usepackage{cite}
\usepackage{amsfonts}
\usepackage{set space}
\usepackage{graphicx}
\usepackage{mathptmx}
\usepackage{latexsym}%
\usepackage{amsmath}
\usepackage{amsthm}

\title{
Consideration on the relationship of theoretical ultraviolet cut-off energy with experimental data
} 
\author{
Kimichika Fukushima
\thanks{E-mail: km.fukushima@mx2.ttcn.ne.jp 
Phone: +81-90-4602-0490 Phone/Fax: +81-45-831-8881}\\
Theoretical Division, South Konandai Science Research,\\
9-32-2-701, Konandai, Konan-ku, Yokohama 234-0054, Japan\\
 \\
Hikaru Sato\\
Emeritus, Department of Physics, Hyogo University of Education,\\
Yashiro-cho, Kato-shi, Hyogo 673-1494, Japan
}

\date{             }

\begin{document}
\setlength{\baselineskip}{18pt}

\maketitle

We formulated a field theory in space-time elements, which are obtained by dividing the space-time continuum into arbitrary-shaped space-time elements, such as hyper-octahedra (orthoplexes), which are aligned locally periodically, without long-range order. Each field is expressed in terms of scalar expansion functions with rotationally covariant coefficients. The cosmological constant was related to the cut-off energy of the quadratic Higgs self-energy. From the literature, examinations find an experimental break in the cosmic ray energy spectrum called 'knee' at about 3 [PeV], which agrees with this theoretical ultraviolet cut-off energy. Moreover, the expansion functions are updated to plane waves using the Lorentz/Poincar\'e covariant/invariant inner product of vectors with each four-momentum within the cut-off energy. Due to the long-range disordered alignment of space-time elements, this study newly introduces an effective field, which is not in the standard model and lies within the Planck energy. Particles are then rarely excited to the energy region between the cut-off energy and the Planck energy via extremely weak interaction solely with the effective field. The theoretical ultraviolet cut-off energy is consistent with the experimental phenomena such as the stability of our Universe, inflation in the early Universe and no detection of neutrinos above PeV-order energy.

\fussy%
\sloppy%

\section{Introduction and Summary}\label{sec:Sec1}

In current theoretical physics and cosmology, some of the
unsolved fundamental problems include
(i) the construction of a consistent field theory without ultraviolet
divergences;
(ii) the identification of dark matter;
(iii)
revealing the origin of the cosmological constant; 
and 
(iv)
the removal of the quadratic divergence of Higgs self-energy.
A Poincar\'e covariant field theory without ultraviolet divergences was formulated
in the previous papers
\cite{Fuku84,Fuku14,Fuku16,Fuku17,Fuku18}
by dividing the space-time continuum into arbitrary-shaped space-time elements like hyper-octahedra,
with straight lines and flat surfaces that are deformed into curves
and curved surfaces containing a space-like hyper-surface, or orthoplexes.
This formalism \cite{Fuku84,Fuku17} is based on the finite element method
\cite{FemMW},
which is widely used and well-posed in the continuum limit,
like other approach \cite{BMS83}.
Each orthoplex in the Minkowski space-time continuum or its Euclidean Wick rotation is mapped to a hyper-cube in a parameter space \cite{Fuku84,Fuku17}, and this one-to-one map has a non-vanishing Jacobian due to the non-existence of point particles.
Fields are expanded in terms of basis functions, which are scalar
\cite{Fuku84,Fuku17}
and are multiplied by an individual coefficient that is a
scalar, vector or tensor depending on the corresponding field type.
(In this paper, basis functions are updated to plane waves with a cut-off.)
At present,
the division of the space-time continuum is not a technical tool, but
is regarded as a fundamental principle, with the ultraviolet cut-off energy as a fundamental constant.
It will depend on the stage of science development
as to whether this principle can be derived from a more fundamental theory beyond the cut-off energy.
\par
To ensure the consistency of general relativity with field theory, the subtracted mass term, with divergent ultraviolet properties in the renormalization
\cite{Weiss,Weiss39,Tomo46,Tomo48,Schwin,Feyn,Dys},
is included into the cosmological constant $\Lambda$
\cite{Eins,Fried}.
Since it was shown in
Ref. \cite{Fuku18}
that zero-point energies do not exist in the expanding universe due
to a lack of periodicity at boundaries, the cosmological constant $\Lambda$ in the present
formalism contains only the self-energy predominantly contributed by the quadratic Higgs term.
The expanding universe treated in this paper has the boundary condition that fields do not exist outside of the expanding universe.
Consequently, the Higgs self-energy, containing the cut-off energy, determines the ultraviolet cut-off energy (and length).
The present formalism thus gives a possible solution to the current
problems
mentioned at the beginning of this section.
Additionally,
the dark matter in the problem (ii)
may be
partially the chromodynamical vacuum, as described in
Ref. \cite{Fuku18}.
The ultraviolet cut-off energy, which is determined theoretically from the present
theory by mapping the theory to the continuum theory with the corresponding cut-off energy, is in a regime around
$3$ [PeV].
Meanwhile,
this theoretical
cut-off energy
of the present theory is much higher than experimental accelerator energies, which are
about $10^4$ [GeV] at present.
In the next stage, the authors compare this
predicted
cut-off energy
with experimental data.
\par
This paper therefore aims to explore a possible relation
between the relativistic cut-off theory
and the features of the cosmic-ray energy spectrum.
The ultraviolet cut-off energy is first examined in detail,
with the help of current particle data
and the corresponding cosmological data.
This study theoretically predicts that
the ultraviolet cut-off energy is
about $3$ [PeV].
Meanwhile, it is reported in
the literature that the experimentally observed energy spectrum of the cosmic-ray flux
shows a break, called
a 'knee', at a corresponding energy of about 3 [PeV].
Thus, cosmic rays decay in the form of a power law
$(E_{\rm c})^{-s}$ with $s$
being the spectral index as the cosmic-ray energy $E_{\rm c}$ increases, showing a greater decline above the knee energy.
It is found that this theoretical cut-off energy
coincides with the experimental value for the knee energy.
This coincidence forms the first finding of this study.
\par
The origin of the cosmic rays with energies of less than about
$1 \times 10^{15}$ [eV]
is understood from astrophysical studies.
In this paper, the basis functions are updated to plane waves with a cut-off,
as described below and Sections \ref{sec:Sec2} and \ref{sec:Sec3}.
This formalism is
Poincar\'e covariant/invariant,
because the formulation is presented using the inner product of vectors, as mentioned
in the paragraphs including Eqs.~(\ref{eqn:vecto})-(\ref{eqn:loint}).
Each momentum lies within the cut-off energy in the normal process.
Additionally,
the cosmic-ray spectrum above the cut-off (knee) energy about 3 [PeV] in the present theory is understood as mentioned below. Similarly to solid state physics, the long-range disorder of the local periodic alignment of space-time elements causes the scattering of the wave function. The effective field that causes this scattering is newly introduced in this paper. This effective field is an Abelian gauge field, which is not contained in the standard model.
Due to the local periodicity of the effective field, as mentioned around
Eqs. (\ref{eqn:perio})-(\ref{eqn:defgn}) and (\ref{eqn:4peep}),
the four-momentum of this effective is discrete and its energy is between the cut-off energy and the Planck energy when the energy is positive. Then,
the highly excited process of a quantum particle
rarely occurs beyond the cut-off energy similarly to solid-state physics,
via the exchange of the above discrete momentum in
an extremely weak interaction solely with the effective field.
In this case, the momentum becomes the sum of two momentums:
one is within the cut-off energy and the other is discrete.
The latter is in the energy region between this
cut-off energy and the Planck energy. 
The above theory is formulated in the tangent space of the 4D space-time continuum.
\par
Furthermore, the theoretical ultraviolet cut-off for fields is consistent with other fundamental astrophysical phenomena, as mentioned in Sec. \ref{sec:Sec3}. The relation between the cosmological constant and the cut-off energy coincides with the stability of our Universe and inflation in the early Universe,
as described in Sec. \ref{sec:Sec3}.
These are also authors' research findings.
Since the interaction between neutrinos and matter is weak, the neutrino shows fundamental physical properties, as is claimed in Ref.
\cite{WenYin}.
The theory of this paper indicates that a light quantum particle is largely scattered above the cut-off energy. This result is
consistent with the experimental data, in which neutrinos and gamma rays are not observed at energies above approximately the knee energy
\cite{Aartsen,Yoshida}.
\par
This article is organized as follows.
Section \ref{sec:Sec2} describes the theoretical ultraviolet cut-off energy.
Section
\ref{sec:Sec3}
then compares the theoretical cut-off energy with experimental cosmic ray data.
It is also shown that particles rarely excited into the energy region between
the cut-off energy and the Planck energy, via an interaction with the
effective filed.
It is then described that the theoretical ultraviolet cut-off in this paper is consistent with the astrophysical phenomena,
including the non-detection of neutrinos presented in the Appendix.
Section \ref{sec:Sec4}
summarizes conclusions.
\par

\section{Formalism of the Ultraviolet Cut-Off Energy}\label{sec:Sec2}

The relationship of the formalism with the theory of relativity is described, before reporting this study.
The theory for the space-time continuum divided into the elements is already defined as a theory for the continuum.
Although a limiting process of $1/\Delta^{(\rm E)}_{\rm M} \rightarrow 0$ for the cut-off energy $\Delta^{(\rm E)}_{\rm M}$ in the fundamental theory may not be essential, the following procedure is a mathematical example of such a process.
The action functional is presented using integrals and differentiations that are defined using an infinitesimal interval $\Delta_{\rm I, D}$. When $1/\Delta^{(\rm E)}_{\rm M} \leq \Delta_{\rm I, D}$, the limiting process $\Delta_{\rm I, D} \rightarrow 0$ does not depend on the shape of the space-time elements.
Furthermore, the
present formalism derives physical quantities using the path integral of the action functional.
When the derived expectation value converges to a unique value within this mathematical framework, this derivation generates validity. Moreover, it is noted that integral and differential manipulations defined in terms of infinitesimal finite differences or measures also require convergence properties.
The relativistic consistency required in this study comprises the following three items:
(i) In the tangent space of the space-time continuum, the formalism is described in terms of Lorentz covariant/invariant functions of the scalar, vector, or tensor.
(ii) Integrals in the formalism are invariant under translation with respect to the coordinates of the system. This symmetry is the Poincar\'e invariance, which is required for the present formalism in the integral form. (iii)
The relativistic cutoff formalism leads to the low-energy form in the limit of the high-energy or small-length cutoff, including gravity.
\par
The physical quantities consistent with the theory of relativity are such as scalars, vectors, and tensors. The vector is determined from the orientation and length in the space-time continuum.
The time and space coordinates are denoted as
$x=x^{\mu}=(x^0,x^1,x^2,x^3)$
$=(ct,{ \bf x})$
$=(ct,x,y,z)$,
with $c$ being the velocity of light \cite{BJDR}.
The inner product between vectors $x=x^{\mu}$ and $y=y^{\mu}$ is defined as
$x \cdot y$$=(g_{\mu\nu}x^{\mu})y^{\nu}$$=x_{\nu}y^{\nu}$$=x^{\mu}(g_{\mu\nu}y^{\nu})$$=x^{\mu}y_{\mu}$.
Here,
$g_{\mu\nu}$ is the metric tensor
and a special form for Minkowski space is given by
\begin{eqnarray}
g_{\mu \nu}=
\left[\begin{array}{cccc}
 1 & 0 & 0 & 0 \\
 0 & -1 & 0 & 0 \\
 0 & 0 & -1 & 0 \\
 0 & 0 & 0 & -1 \\
\end{array}\right]
\label{eqn:metr0},
\end{eqnarray}
using the notation defined in
\cite{BJDR}. 
For a vector $x_{(2)}-x_{(1)}$, the Lorentz transformation represented by the matrix $L_{\mu}^{\nu}$ is given by
\begin{eqnarray}
x_{(2)}-x_{(1)}
=x^{\nu}_{(2)}-x^{\nu}_{(1)}
=L^{\nu}_{\mu}(x^{\prime \mu}_{(2)}-x^{\prime \mu}_{(1)})
\label{eqn:vecto},
\end{eqnarray}
yielding the following squared vector:
\begin{eqnarray}
\hspace{-5ex}
(x_{(2)}-x_{(1)})
\cdot
(x_{(2)}-x_{(1)})
&=&
(x^{\nu}_{(2)}-x^{\nu}_{(1)})
(x_{\nu (2)}-x_{\nu (1)})
\nonumber\\
\hspace{-5ex}
=L^{\nu}_{\mu}(x^{\prime \mu}_{(2)}-x^{\prime \mu}_{(1)})
 L_{\nu}^{\eta}(x^{\prime}_{\eta(2)}-x^{\prime}_{\eta(1)})
&=&
\delta_{\mu}^{\eta}(x^{\prime \mu}_{(2)}-x^{\prime \mu}_{(1)})
(x^{\prime}_{\eta(2)}-x^{\prime}_{\eta(1)})
\nonumber\\
\hspace{-5ex}
&=&
(x^{\prime}_{(2)}-x^{\prime}_{(1)})
\cdot
(x^{\prime}_{(2)}-x^{\prime}_{(1)})
\label{eqn:vescto},
\end{eqnarray}
where the Kronecker delta $\delta_{\mu}^{\eta}$ has been used.
The
vector $x_{(2)}-x_{(1)}$ is the relative quantity,
and the squared vector $(x_{(2)}-x_{(1)})\cdot(x_{(2)}-x_{(1)})$ is the
Lorentz
invariant squared length.
Furthermore, for the translation of coordinates by the functions  $x^{\mu}_{(1)}=x^{\prime \mu}_{(1)}+a^{\mu}$ and $x^{\mu}_{(2)}=x^{\prime \mu}_{(2)}+a^{\mu}$, the vector
$x^{\mu}_{(2)}-x^{\mu}_{(1)}=x^{\prime \mu}_{(2)}-x^{\prime \mu}_{(1)}$ is covariant. These are the covariance/invariance of vectors for the Lorentz and Poincar\'e transformations.
\par
In the case of the integral of the scalar wave function
\begin{eqnarray}
\hspace{-2ex}
S=\int dx^0dx^1dx^2d^3
\phi_k(x),
\hspace{2ex}
\mbox{with}
\hspace{1ex}
\phi_k(x)
=\exp(ik_{\mu} x^{\mu})
\label{eqn:action},
\end{eqnarray}
the quantity $x^{\mu}$ can be regarded as a vector $x_{(2)}-x_{(1)}$ for $x^{\mu}_{(2)}=x^{\mu}$ and $x^{\mu}_{(1)}=(0,0,0,0)$, that is,
the coordinate origin. Using Eq.~(\ref{eqn:vecto}) and
\begin{eqnarray}
\int dx^0dx^1dx^2d
x^3
=
\int
dx^{\prime 0}
dx^{\prime 1}
dx^{\prime 2}
dx^{\prime 3}
\label{eqn:loint},
\end{eqnarray}
the integral in Eq.~(\ref{eqn:action}) is covariant/invariant for the
Poincar\'e transformation.
The above relations are examples of the covariance/invariance pertaining to the
Lorentz/Poincar\'e transformation of in this paper.
In general, the tangent space on the curved space-time continuum enables the simpler local treatment of the Lorentz/Poincar\'e covariance/invariance.
The quantity $\sqrt{{\rm \det}(g_{\mu\nu})}$ that appears in general relativity corresponds to the Jacobian for integral with respect to the coordinate variables at each point on the curved space-time continuum.  At each point, general relativity requires the existence of the tangent space that is flat with the metric defined using Eq.~(\ref{eqn:metr0}). Hence, fields can be defined on such a tangent space.
\par
Among electromagnetic, weak and strong interactions, the leading self-energy is the quadratic Higgs term.
We are concerned with the relation between this self-energy and general relativity, described by the following Einstein field equations \cite{LanL} with $\mu, \nu=0,1,2,3$
\begin{eqnarray}
R_{\mu\nu}-\frac{1}{2}g_{\mu\nu}R=\frac{8\pi G}{c^4} T_{\mu\nu}
\label{eqn:Eeq}.
\end{eqnarray}
Here, $G$ is the gravitational constant;
and
$T_{\mu\nu}$ is the energy-momentum tensor.
In Eq.~(\ref{eqn:Eeq}), the scalar curvature $R$ is sequentially defined from the Riemann curvature tensor $R^{\rho}_{\mu \rho \nu}$ as
follows:
$
R_{\mu \nu}=R^{\rho}_{\mu \rho \nu},
$
and
$
R=g^{\mu \nu} R_{\mu \nu}.
$
The notations  used for the indices in
these contractions
as well as the expressions of the Einstein equations follow those employed in
\cite{LanL}. In this textbook, it is noted that in the literature,
a definition of the contraction using another pair of
indices yields a difference in sign.
\par
We expressed the following
mass term with the ultraviolet divergence, and in the renormalization this term is subtracted from the Hamiltonian in the form
\begin{eqnarray}
T^{(\rm S)}_{\mu\nu}=\frac{c^4}{8\pi G}g_{\mu\nu}\Lambda^{(\rm S)}
\label{eqn:Smtm}.
\end{eqnarray}
The Einstein field equations that determine the space-time curvature strongly require that all energies be included. The energy subtracted in the renormalization for fields is not an exception.
To ensure the consistency of general relativity with field theory, the subtracted mass term
given by
Eq.~(\ref{eqn:Smtm}) is explicitly expressed in
the Einstein field equations in
Eq.~(\ref{eqn:Eeq}), giving
\begin{eqnarray}
R_{\mu\nu}-\frac{1}{2}g_{\mu\nu}R=\frac{8\pi G}{c^4}
\left(
T_{\mu\nu}-T^{(\rm S)}_{\mu\nu}
\right)
\label{eqn:EeqR}.
\end{eqnarray}
Using Eq.~(\ref{eqn:Smtm}), the above equations are rewritten as
\begin{eqnarray}
R_{\mu\nu}-\frac{1}{2}g_{\mu\nu}R+ g_{\mu\nu} \Lambda=\frac{8\pi G}{c^4}
T_{\mu\nu}
\label{eqn:EeqLam},
\end{eqnarray}
where
$\Lambda=$
$\Lambda^{(\rm S)}$
in Eqs.
(\ref{eqn:Smtm})
and (\ref{eqn:EeqLam})
has been be regarded as the cosmological (vacuum energy) constant $\Lambda$.
\par
In Ref.
\cite{Fuku18},
the cosmological constant was related to the quadratic Higgs self-energy, which is larger than the other self-interactions.
In the present theory, gravity is weak at the cut-off energy compared to the Planck energy.
In Ref.
\cite{Fuku18},
it was shown that zero-point energies cannot exist, due to the inconsistency with a lack of periodicity at the boundaries of the expanding universe.
Since the main component of baryonic matter is proton in hydrogen, the dominant contributions to the Higgs self-energy are loops of fermions (up and down quarks) interacting with the Higgs field.
In essence, the left side of the Einstein equations in Eq.~(\ref{eqn:Eeq}) describes gravitational fields, while the energy-momentum tensor on the right side gives the mass (energy) term that is the source of the gravitational field.
The enhanced mass of the Higgs field due to the self-interaction for the coupling constant is calculated using quantum field theory, and the derived expectation value is presented on the right side. A term representing the Higgs self-energy multiplied by $-1$ is added to the right side of the Einstein field equations, not after but during the renormalization.
(This negative mass term is moved to the left side.)
\par
The squares of the self-energy of the Higgs field in natural units with $\hbar=1, c=1$, where $\hbar$ is the reduced Planck constant and $c$ is the velocity of light, is written by
\begin{eqnarray}
\vert E_{\rm H}\vert ^2=\frac{1}{8\pi^2}
(\vert \lambda_{\rm u}\vert ^2+\vert \lambda_{\rm u}\vert ^2+\vert \lambda_{\rm d}\vert ^2)
\left(
\Delta^{(\rm P)}_{\rm E}
\right)^2
\label{eqn:SEHg},
\end{eqnarray}
where
$\Delta^{(\rm P)}_{\rm E}$
is the Euclidean ultraviolet cut-off
parameter, implying the length of the four-vector, one component of which is the energy.
The quantities
$\lambda_{\rm u}$ and $\lambda_{\rm d}$ are the coupling constants of the Higgs field with up
and down quarks, respectively, given by
$
\lambda_{\rm u}=
(\sqrt{2}/{v})m_{\rm u}
$
and
$
\lambda_{\rm d}=
(\sqrt{2}/{v})m_{\rm d}.
$
Here, $m_{\rm u}$ and $m_{\rm d}$ are
the masses of the up and down quarks, respectively, and
$v$ is the vacuum expectation value of the
Higgs field after symmetry breaking.
The quantities $v$, $\lambda_{\rm u}$, $\lambda_{\rm d}$, $m_{\rm u}$, and $m_{\rm d}$
appear in the density of Lagrangian that denotes the Yukawa interaction between the Higgs field and quarks.
A detailed derivation of the dominant Higgs self-energy is given in Ref.
\cite{higgsse},
and can be summarized as follows.
The fermion loop contribution to the enhancement of the Higgs mass depicted by the Feynman diagram in the up quark case with the index ${\rm u}$ is written as
\begin{eqnarray}
\vert E_{\rm H u}\vert ^2=i\int \frac{d k^4}{(2\pi)^4}
{\rm Tr}\left[\frac{-i\lambda_{\rm u}}{\sqrt{2}}
\frac{i}{\slashed{k} -m_{\rm u}}
\left(\frac{-i\lambda^*_{\rm u}}{\sqrt{2}}\right)
\frac{i}
{\slashed{k} -m_{\rm u}}
\right],
\label{eq:higlp}
\end{eqnarray}
where
$\slashed{k}=\gamma^{\mu}k_{\mu}$ with $\gamma^{\mu}$ being Dirac's gamma matrices.
The Wick rotation by $k_0 \rightarrow ik_4$ and $k^2 \rightarrow -(k_{\rm E})^2$, where
$k_{\rm E}$
is the Euclidean momentum, yields
\begin{eqnarray}
\vert E_{\rm H u}\vert ^2=-\frac{\vert \lambda_{\rm u} \vert ^2}{8 \pi^2}
\int_0^{(\Delta^{(\rm P)}_{\rm E})^2}
d k_{\rm E}^2\frac{(k_{\rm E})^2[(k_{\rm E})^2-(m_{\rm u})^2]}
{(k_{\rm E})^2-(m_{\rm u})^2}
\label{eq:wrhip}.
\end{eqnarray}
In the Minkowski space, this approximates to
\begin{eqnarray}
\vert E_{\rm H u}\vert ^2=
\frac{1}{8\pi^2}
\vert \lambda_{\rm u}\vert ^2
(\Delta^{(\rm P)}_{\rm E})^2
\label{eq:mmeh}.
\end{eqnarray}
\par
The dominant element of our Universe is hydrogen, which contains the proton. The matter energy is then due to the proton energy. The Yukawa coupling between the quarks and Higgs boson is expressed in the Lagrangian density formalism, and the mass energy arises from the interaction via the following
Higgs field $v$ in the expression
$
\lambda_{\rm u}=
(\sqrt{2}/{v})m_{\rm u}
$
in the proton. The vacuum energy, which is the self-energy of the Higgs boson, is due to this
Higgs boson, and is measured by the proton energy.
The vacuum energy, which is measured by the proton energy,
is expressed in terms of the cosmological constant $\Lambda$ as
\begin{eqnarray}
E_{\Lambda}=
\frac{\Omega_{\Lambda}}{\Omega_{\rm m}}
m_{\rm p}
\label{eqn:EVauL},
\end{eqnarray}
where
$m_{\rm p}$
is the proton mass energy, and
$\Omega_{\rm m}$
and
$\Omega_{\Lambda}$
are the fractions of matter energy and vacuum energy (expressed
by the cosmological constant
$\Lambda$) in
the total energy of the universe, respectively.
Since the Lagrangian density is used in the present formalism,
$\Omega_{\Lambda}/\Omega_{\rm m}$ in Eq. (\ref{eqn:EVauL}) is also equal to the ratio of the vacuum-energy density to the matter density.
Symbols such as $\Omega_{\rm m}$
are fractions of the total energy of the universe in this paper,
whereas usually they are fractions of the critical density. Since the former
quantities
have the same values as the latter, identical symbols are used for
both in this paper.
It is well known that the masses of the three quarks composing a proton with
mass
$m_{\rm p}$
are quite small. Using $m_{\rm u}$ and $m_{\rm d}$,
the following ratio can be defined:
\begin{eqnarray}
\gamma_{\rm P}=\frac{
m_{\rm p}
}{m_{\rm u}+m_{\rm u}+m_{\rm d}}
\label{eqn:Pfra}.
\end{eqnarray}
Then, Eq.~(\ref{eqn:EVauL}) is rewritten as
\begin{eqnarray}
E_{\Lambda}=
\frac{\Omega_{\Lambda}}{\Omega_{\rm m}}
\gamma_{\rm P}(m_{\rm u}+m_{\rm u}+m_{\rm d})
\label{eqn:EVPfra}.
\end{eqnarray}
By equating $E_{\Lambda}$ in Eq.~(\ref{eqn:EVPfra}) to $E_{\rm H}$ in Eq.~(\ref{eqn:SEHg}) and using
Eqs.~(\ref{eq:higlp})$
-$(\ref{eqn:Pfra}),
it follows that
\begin{eqnarray}
\hspace{-2ex}
\frac{1}{2\pi v}
(m_{\rm u}^2+m_{\rm u}^2+m_{\rm d}^2)^{1/2}
\Delta^{(\rm P)}_{\rm E}
=\frac{\Omega_{\Lambda}}{\Omega_{\rm m}}
\gamma_{\rm P}(m_{\rm u}+m_{\rm u}+m_{\rm d})
\label{eqn:EqEHL},
\end{eqnarray}
and we derive the following theoretical Euclidean cut-off parameter
\begin{eqnarray}
\Delta^{(\rm P)}_{\rm E}
=
(2\pi v)
\frac{\Omega_{\Lambda}}{\Omega_{\rm m}}
\gamma_{\rm P}
\frac{(m_{\rm u}+m_{\rm u}+m_{\rm d})}
{(m_{\rm u}^2+m_{\rm u}^2+m_{\rm d}^2)^{1/2}}
\label{eqn:ECO}.
\end{eqnarray}
Using the experimental data presented in the following section, calculations show that the subtracted divergent part contained in the self-energy of the Higgs boson and included in the cosmological constant in the renormalization is approximately 13.5 [GeV]. Whereas, the experimental mass of the Higgs boson is approximately 125-126 [GeV].
A description of the theoretical origin of the cosmological constant is seen in Ref.
\cite{Birrell}.
\par
The above ultraviolet cut-off energy is generally used in Euclidean space,
and is obtained from an
analytic continuation with Wick rotation from Minkowski space. The energy in Minkowski space $E_{\rm M}$ is replaced by $E_{\rm E}$, and the absolute value of the cut-off
parameter
$
\vert \Delta^{(\rm P)}_{\rm E}\vert 
$
in Euclidean space may be equal to the cut-off energy $\Delta^{(\rm E)}_{\rm M}$ in Minkowski space.
Despite this expectation value for the cut-off energy,
the following possibility is deduced for the ultraviolet cut-off energy in Minkowski space. The energy in Minkowski space is written as
\begin{eqnarray}
E_{\rm M}=[(m_0)^2+\vert {\bf p}\vert ^2]^{(1/2)}
\label{eqn:EMin},
\end{eqnarray}
where $m_0$ is the rest
mass and
${\bf p}$ is the
three-dimensional momentum.
Meanwhile, the integral over an infinitesimal volume with respect to the Euclidean four-vector $p_{\rm E}$ is written as:
$
d^4 p_{\rm E}=\pi^2  (p_{\rm E})^2 d^2 p_{\rm E}.
$
Considering the effective equivalence for the transformation $-(p_{\rm E})^2 \rightarrow (p_{\rm E})^2$ in the above integral,
the squared length of the Euclidean four-vector becomes:
\begin{eqnarray}
\vert (p_{\rm E})^2\vert 
=(E_{\rm M})^2+{\vert \bf p}\vert ^2=(m_0)^2+2\vert {\bf p}\vert ^2
\label{eqn:EEuc}.
\end{eqnarray}
For large kinetic energies with corresponding large
$\vert {\bf p}\vert $,
this equation is approximated as
$
\vert (p_{\rm E})^2\vert 
\approx 2(E_{\rm M})^2,
$
yielding
$
E_{\rm M} \approx
(1/\sqrt{2})
\vert p_{\rm E}\vert.
$
We then set
$
\Delta^{(\rm E)}_{\rm M}=(1/\sqrt{2})
\Delta^{(\rm P)}_{\rm E},
$
where $\Delta^{(\rm E)}_{\rm M}$ is
the ultraviolet cut-off energy for Minkowski space, and
$\Delta^{(\rm P)}_{\rm E}$ is the cut-off parameter in Eq.~(\ref{eqn:SEHg}) for
Euclidean space.
The symbol $\Delta^{(\rm P)}_{\rm E}$ is not a cut-off energy for Euclidean space but the upper bound of the integral in
Eq.~(\ref{eq:wrhip}).
\par

\section{Comparison of the Theoretical Ultraviolet Cut-Off Energy and Cosmic Ray Spectrum with Experimental Data}\label{sec:Sec3}

Here,
the relativistic cut-off
energy is 
compared with experimental spectra.
The data used were
$\Omega_{\Lambda}$
$=0.694$, 
$\Omega_{\rm b}$
$=0.0483$, $m_{\rm u}=2.2$ [MeV],
$m_{\rm d}=4.7$ [MeV], 
$m_{\rm p}$
$=938.272$ [MeV],
$v=246.22$ [GeV] 
\cite{Tanabashi18,Arason}, where 
$\Omega_{\rm b}$
is the energy fraction of baryons of the total universe energy.
The relativistic cut-off theory was then used to derive
the theoretical Minkowski cut-off energy $\Delta^{(\rm E)}_{\rm M}=(1/\sqrt{2})\Delta^{(\rm E)}_{\rm E} (=7.5 \times 10^{-8} \mbox{ [fm]})
=2.6\mbox{ [PeV]} \approx 3 \mbox{ [PeV]}$.
Meanwhile, experimental cosmic rays, which are thought to be produced by remnants of a supernova (star explosion),
have the following energy spectrum characteristics. At low energies, the
energy spectrum of cosmic rays rises with increasing cosmic ray energy,
and has a peak at the energy $E_{0} \approx 10^8$ [eV]
followed by a decline in the form of $(E_{\rm c})^{-s}$, where $E_{\rm c}$ is the cosmic-ray energy and $s$ is the spectral index.
The spectrum has the form $\approx E_{\rm c}^{-2.7}$ for $E_{\rm c} \leq 3 \times 10^{15}$ [eV],
and changes to the form $\approx E_{\rm c}^{-3.0}$ for $E_{\rm c} \geq 3 \times 10^{15}$ [eV].
Hence, a spectrum break occurs at an energy $E_{\rm c} \approx 3 \times 10^{15}$ [eV], and this break is called the knee.
(The spectrum for energies of less than the energy of about $10^{15}$ [eV] is understood based on astrophysical studies.)
Thus, the theoretical ultraviolet cut-off energy of about
$3$ [PeV]
derived coincides with the experimentally observed knee energy of about
3 [PeV].
There is therefore a possibility that the ultraviolet cutoff exists
at the knee energy of 3 [PeV],
as specified from the ultraviolet cut-off energy.
\par
Let us now consider the relation between the properties of the present
cut-off formalism mentioned below and those of the experimental energy spectrum.
In solid state physics
\cite{Kittel56,Kittel63},
the wave function of an electron is a function of each point of the space continuum.
Unit cells are obtained by dividing a crystal and are aligned in this space continuum.
For instance, a cubic cell has an atom at its center, which is separated by distance $a_{\rm s}$ from the atom of the nearest neighbor. 
Then, the static wave function in one dimension (1D)
is restricted to the following form of the Bloch function
\begin{eqnarray}
\phi_{{\rm s},k_{\rm s}}(x) =
[\exp(ik_{s}x)]
u_{\rm s}(x)
\label{eqn:Bloch},
\end{eqnarray}
where $x$ is a point in the space continuum and $k_{\rm s}$ is the momentum,
which is restricted
to $-\pi/a_{\rm s} \leq k_{\rm s} \leq \pi/a_{\rm s}$.
The function $u_{\rm s}(x)$ is periodic and satisfies
\begin{eqnarray}
u_{\rm s}(x+a_{\rm s})=u_{\rm s}(x)
\label{eqn:perio}.
\end{eqnarray}
As is written in the literature
\cite{Kittel63},
using discrete reciprocal lattice vectors
$
G_{n_{\rm s}}=n_{\rm s}(2\pi/a_{\rm s})
$
that satisfies
$\exp(-iG_{n_{\rm s}} a_{\rm s})=1$
$
\mbox{with $n_{\rm s}$ being integers},
$
this
$u_{\rm s}(x)$ can be expanded as
\begin{eqnarray}
u_{\rm s}(x)
=\sum_{n_{\rm s}} u_{{\rm s} n_{\rm s}} \exp(iG_{n_{\rm s}}x).
\label{eqn:1drex}
\end{eqnarray}
When the momentum $P_{\rm s}=k_{\rm s}+G_{n_{\rm s}}$
from Eqs. (\ref{eqn:Bloch}) and (\ref{eqn:1drex})
is restricted to $\vert P_{\rm s} \vert \leq \pi/a_{\rm s}$, the quantity 
$\pi/a_{\rm s}$ is
a cut-off momentum.
\par
In the 4D case, the Euclidean/Minkowski space-time continuum in the tangent space is divided into arbitrary-shaped elements containing space-like hypersurfaces. The centers of the nearest-neighbor space-time elements are separated by $a_r$. The alignment of these space-time elements is locally periodic, and there is no long-distance order.
The center of the space-time element is denoted as
$x_p=x^{\mu}_p=(x^0_{p},x^1_{p},x^2_{p},x^3_{p})$, and
$G_{\tilde{n} p}=$
$G_{\mu \tilde{n} p}$
in notation at this point are defined as
\begin{eqnarray}
G_{0 \tilde{n} p}
&=&
n_0(2\pi/a_r)e_{0p},
\hspace{2ex}
G_{1 \tilde{n} p}=n_1(2\pi/a_r)e_{1p},
\nonumber\\
G_{2 \tilde{n} p}
&=&
n_3(2\pi/a_r)e_{2p},
\hspace{2ex}
G_{3 \tilde{n} p}=n_3(2\pi/a_r)e_{3p}
\label{eqn:defgn},
\end{eqnarray}
where
$n_0, n_1, n_2, n_3$
are integers.
The above $e_{\mu p}$ at the point $x_p=x^{\mu}_p$ are unit vectors oriented in individual arbitrary directions under the condition that they are orthogonal to each other
$
e_{\xi \mu p}e^{\xi}_{\nu p}=\delta_{\mu\nu}.
$
These unit vectors, $e_{\mu p}$, specify local periodic directions of the alignment of space-time elements.
Let $k_{\mu}$ be the momentum
of a quantum particle in the first momentum zone expressed as
\begin{eqnarray}
\vert k\vert =\vert k_{\mu}\vert =\vert (k_{\mu}k^{\mu})^{1/2}\vert  \leq \pi/a_r
\label{eqn:wakmuq}.
\end{eqnarray}
The momentum $P_{\mu}$
of the quantum particle becomes
$P_{\mu}
=k_{\mu}+G_{\mu \tilde{n} p},
$
and its associated wave function of the particle has the following form:
\begin{eqnarray}
\psi_P(x)
=
(1/c_{\rm N})u_P\exp(iP_{\mu}x^{\mu})
\label{eqn:wa4pq},
\end{eqnarray}
where $c_{\rm N}$ is a normalization constant.
\par
Here, this study explores the effect of the disordered alignment of space-time elements.
Let us consider an arbitrary-shaped space-time element centered at point $x_{p}=x^{\mu}_p$. Around this point, the local periodic alignment of space-time elements is oriented along arbitrary directions, which are denoted by unit vectors $e_{\mu p}$
of reciprocal vectors $G_{\mu \tilde{n} p}$
in Eq.~(\ref{eqn:defgn}).
When a quantum particle
initially at at the point $x_{p}=x^{\mu}_p$
passes through a space-time element centered at $x_{q}=x^{\mu}_q$ with arbitrary local periodic directions denoted by vectors $e_{\mu q}$
of $G_{\mu \tilde{n} q}$,
the directions $e_{\mu p}$ of $G_{\mu \tilde{n} p}$ are different from $e_{\mu q}$
of $G_{\mu \tilde{n} q}$
at $x_{q}$.
The wave function
of the particle is scattered due to this local order (long-range disorder), similarly to solid state physics.
Then, the following disorder-induced effective field arises due to the change of
$G_{\mu \tilde{n} p}$ to $G_{\mu \tilde{n}^{\prime} q}$
to change the momentum of the particle
\begin{eqnarray}
A_{\mu{\rm E}}(x)
=
\sum_p
\sum_{\tilde{n}} A_{ {\mu{\rm E}} {\tilde{n}} } \exp(iG_{\nu \tilde{n} p}x^{\nu})
\label{eqn:4peep}.
\end{eqnarray}
(This disorder-induced effective field can be second-quantized, and neutrino scatterings by this field are described in Appendix.)
It is noted that energies of momentums $G_{\mu \tilde{n} p}$, which are higher or equal to the ultraviolet cut-off energy $\Delta^{(\rm E)}_{\rm M}$ about 3 [PeV] for $\vert G_{\mu \tilde{n} p} \vert >0$, are set to values below the Planck energy.
The Hamiltonian density of a particle in the standard model is expressed as
\begin{eqnarray}
{\cal H}={\cal H}_{0} +{\cal H}_{\rm I}+{\cal H}_{\rm IE}
\label{eqn:hamsp},
\end{eqnarray}
where ${\cal H}_0$ is the non-interacting term, and ${\cal H}_{\rm I}$ is the term of interactions with fields in the standard model. As the usual cut-off, the above term ${\cal H}_{\rm I}$ gives rise to normal scattering processes, which forbid the particle from being excited beyond the cut-off energy $\Delta^{(\rm E)}_{\rm M}$
below Eq. (\ref{eqn:EEuc}) derived from Eq. (\ref{eqn:ECO}),
by setting the
coupling constant in ${\cal H}_{\rm I}$
to a value of zero.
On the other hand, the term ${\cal H}_{\rm IE}$ denotes an interaction of the particle in the standard model with the effective field $A_{\mu{\rm E}}$.
Due to this interaction ${\cal H}_{\rm IE}$, a momentum
$G_{\mu {\tilde{n}}^{\prime} q}-$
$G_{\mu \tilde{n} p}$
is exchanged, and the particle is excited beyond the cut-off energy $\Delta^{(\rm E)}_{\rm M}$. Since the value of the fine structure constant $\alpha_{\rm D}$ in this interaction ${\cal H}_{\rm EI}$ is around  $10^{-28}$, as is shown in the Appendix, the transition probability is relatively tiny. In the present scheme, the energy of the particle is further restricted to be less than the Planck energy, which is much lower than the Landau pole energy. The self-energy via the interaction with the effective field $A_{\mu {\rm E}}$
is less than the Higgs self-energy, which is in the standard model, due to the relatively tiny fine structure constant $\alpha_{\rm D}$ and a self-energy (as a logarithmic function of the Planck energy) unlike the Higgs self-energy (as a quadratic function of the cut-off energy $\Delta^{(\rm E)}_{\rm M}$). (As is known generally, a running fine structure constant $\alpha_g(Q^2)$ of a general Abelian gauge field as a function of squared four-dimensional momentum $Q^2$ is given by $\alpha_g(Q^2)=\alpha_g(\mu_g^2)/\{1-[(\alpha_g(\mu_g^2)/(3\pi)]\log(Q^2/\mu_g^2) \}$), where the low energy fine structure constant is defined by $\alpha_g(\mu_g^2)$. The quantity $\alpha_g(Q^2)$ has a singularity at the Landau pole.) Consequently, the energy of particles in the standard model is within the cut-off energy $\Delta^{(\rm E)}_{\rm M}$ in the normal scattering process. These particles are rarely excited beyond the cut-off energy
via the interaction with the effective Abelian field $A_{\mu {\rm E}}$.
Since all energies of particles in the present formalism are less than the Planck energy,
which is much lower than the Landau pole energy,
the present formalism is not affected by the Landau pole.
\par
Here, the coefficients $u_P$
in Eq. (\ref{eqn:wa4pq})
are determined from an initial condition, as mentioned below. Generally, cosmic rays are macroscopic objects, because these rays are possibly created by macroscopic explosions. Field theory of elementary particles is a microscopic theory and the prediction of complex macroscopic phenomena is not always simply and directly obtained. Instead, to describe the cosmic-ray spectrum, one way is to give the coefficients $u_P$ as an initial condition. The initial condition is a given quantity and may contain parameters, which are not
always predicted from the microscopic field theory.
The spectrum then represents a real physical phenomenon and has no additional
wave function structure.	
Let us consider the following two wave packet
$B_{I,t}^{(2)}(t)$ with $I=1, 2$
for time $t$
\begin{eqnarray}
B_{t}^{(2)}(t)=C_1 B_{1,t}^{(2)}(t)+C_2 B_{2,t}^{(2)}(t)
\label{eqn:B212},
\end{eqnarray}
where $C_1$ and $C_2$ are normalized
weighting constants,
such as $C_{1}=C_{2}=1/2$.
If necessary, the
functions $B_{I,t}^{(2)}(t)$ with $I>3$ are added to
the above expression for
$B_{t}^{(2)}(t)$.
The above localized functions $B_{I,t}^{(2)}(t)$ with $I=1,2$ are defined as
\begin{eqnarray}
\hspace{-1ex}
B_{1,t}^{(2)}(t)=
\frac{1}{C_{B}}
\frac{\sqrt{2}}{\Gamma(s_1 /2)}
\left\vert 
\frac{t}{2a_1}
\right\vert 
^{(s_1-1)/2}
K_{(s_1-1)/2}(a_1\vert t\vert )
\label{eqn:B21a},
\end{eqnarray}
\begin{eqnarray}
\hspace{-1ex}
B_{2,t}^{(2)}(t)=
\frac{1}{C_{B}}
\frac{\sqrt{2}}{\Gamma(s_2 /2)}
\left\vert 
\frac{t}{2a_2}
\right\vert 
^{(s_2-1)/2}
K_{(s_2-1)/2}(a_2\vert t\vert )
\label{eqn:B1b2},
\end{eqnarray}
where 
$C_{B}$ is a normalization constant, and $a_1, a_2, s_1$
and
$s_2$ are constants
that give localization properties in the time coordinate
$t$
and corresponding spectrum properties in the coordinate
of the cosmic ray energy
$E_{\rm c}$,
as mentioned below. The function
$K_{\nu}(t)$, in which the
conventional notation $\nu$ in this case refers to
a real number and
$\nu=(s_1-1)/2=s_1/2-1/2$ or $\nu=(s_2-1)/2$,
is the modified Bessel function
\cite{MEHandbook}, 
written as
\begin{eqnarray}
K_{\nu}(t)=\frac{\pi}{2}\frac{I_{-\nu}(t)-I_{\nu}(t)}{\sin(\nu \pi)}
\label{eqn:BesFn}
\hspace{2ex}\mbox{ with }
I_{\nu}(t)=
\left(
\frac{t}{2}
\right)
^{\nu}
\sum_{n=0}^{\infty}
\frac{(t/2)^{2n}}{n!\Gamma(\nu+n+1)},
\end{eqnarray}
where $\Gamma$ is the gamma function.
The function $\vert t\vert ^{\nu}K_{\nu}$ in Eqs.~(\ref{eqn:B21a})
and
(\ref{eqn:B1b2}) has the following localized properties.
Near $t=0$,
\begin{eqnarray}
\vert t\vert ^{\nu}K_{\nu}(t)
\approx
\vert t\vert ^{\nu}\frac{\pi}{2}\frac{1}{\sin(\nu \pi)}
\left(
\frac{t}{2}
\right)
^{-\nu}
\frac{1}{\Gamma(-\nu+1)}
=
\mbox{constant}
\label{eqn:AppI},
\end{eqnarray}
while in the limit as $t \rightarrow \infty$, the function vanishes, leading to
\begin{eqnarray}
\vert t\vert ^{\nu}K_{\nu}(t) \approx \vert t\vert ^{\nu}
\sqrt{\frac{\pi}{2t}}
\exp(-t)
\label{eqn:IIAppb}.
\end{eqnarray}
In Eq.~(\ref{eqn:B212}), the condition
$B^{(2)}_t(0)$
$=
C_1 B_{1,t}^{(2)}(0)+C_2 B_{2,t}^{(2)}(0)=1
$
is imposed
on the absolute square of the localized function $B^{(2)}_t(t)=\vert B_t(t)\vert ^2$,
to simplify the formulation.
Using
Eq.~(\ref{eqn:AppI})
\begin{eqnarray}
B_t^{(2)}(0)
&=&
\frac{C_1}{C_{B}}
\frac{\sqrt{2}}{\Gamma(s_1 /2)}
\lim_{\epsilon\to 0}
\left[
\left\vert 
\frac{\epsilon}{2a_1}
\right\vert 
^{(s_1-1)/2}
K_{(s_1-1)/2}(a_1\vert \epsilon\vert )
\right]
\nonumber\\
&&
\hspace{-1.5ex}
+\frac{C_2}{C_{B}}
\frac{\sqrt{2}}{\Gamma(s_2 /2)}
\lim_{\epsilon\to 0}
\left[
\left\vert 
\frac{\epsilon}{2a_2}
\right\vert 
^{(s_2-1)/2}
K_{(s_2-1)/2}(a_2\vert \epsilon\vert )
\right]
\label{eqn:BCon}.
\end{eqnarray}
By setting
$ C_1=C_2=1/2$, 
the constant $C_B$ is simply obtained from the above Eq.~(\ref{eqn:BCon}).
The below experimental spectrum, which
is consistent with this theoretical spectrum that
is described by two
wave packets in the time coordinate,
suggest that mainly two different types of phenomena such as explosions occurred and generated cosmic rays in our Universe history, as seen in the present analysis.
\par
The functions $B_{I,t}^{(2)}(t)$ with $I=1,2$ are then Fourier-transformed
as follows
\begin{eqnarray}
B_{I,E}^{(2)}(E_{\rm c})
=
\frac{1}{\sqrt{2\pi}}\int_{-\infty}^{\infty} dt\hspace{1ex}
\exp(iE_{\rm c}t) B_{I,t}^{(2)}(t).
\label{eqn:FtBf}
\end{eqnarray}
The above functions 
with I=1,2
and the
$B_{E}^{(2)}$ corresponding to 
$B_{t}^{(2)}$ in Eq.~(\ref{eqn:B212})
can be expressed as
\begin{eqnarray}
\nonumber
B_{1,E}^{(2)}(E_{\rm c})
&=&
\frac{1}{C_{B}}
\frac{1}{[(a_1)^2+(E_{\rm c})^2]^{(s_1 /2)}},\\
B_{2,E}^{(2)}(E_{\rm c})
&=&
\frac{1}{C_{B}}
\frac{1}{[(a_2)^2+E_{\rm c})^2]^{(s_2 /2)}}
\label{eqn:B2E1},
\end{eqnarray}
and
$
B_{E}^{(2)}(E_{\rm c})
=
C_1 B_{1,E}^{(2)}(E_{\rm c})+C_2 B_{2,E}^{(2)}(E_{\rm c}).
$
For the inequalities $a_{1}$
$\ll$
$a_{2}$ and
$s_1 >$
$s_2$,
the functions in the above equation have the form
$B_{E}^{(2)} \approx$
$B_{1,E}^{(2)} \propto (E_{\rm c})^{-s_1}$ for $a_{1} \leq E_{\rm c} \leq a_2$, while
$B_{E}^{(2)} \approx$
$B_{2,E}^{(2)} \propto (E_{\rm c})^{-s_2}$ for $a_{2} \leq  E_{\rm c}$.
(The function
$B_{2,E}^{(2)}$ is small for $E_{\rm c} \leq a_2$
since
$a_1$
$\ll$
$a_2$, while
for
the condition $s_1 >s_2$, the function $B_{1,E}^{(2)}$ falls rapidly compared to $B_{2,E}^{(2)}$ for
$a_2 \leq E_{\rm c}$.)
Here, the constants are set as $a_1 = 3 \times 10^{15}$ [eV]
($\approx \Delta^{(\rm E)}_{\rm M}$, which is the cut-off)
$a_2 = 3 \times 10^{18}$ [eV],
$s_1 = 3.0$ and $s_2 = 2.6$.
The energy spectrum is expressed in terms of the functions in
Eq.~(\ref{eqn:B2E1}),
which may correspond to coefficients $u_P$ in Eq.~(\ref{eqn:wa4pq}).
Then, using this expression, 
the experimentally observed energy spectrum of the cosmic rays is described as
$\approx (E_{\rm c})^{-3.0}$
between the knee energy at about
$3 \times 10^{15}$ [eV]
and the ankle energy at about
$3 \times 10^{18}$ [eV].
Furthermore, this spectrum is also described as
$\approx (E_{\rm c})^{-2.6}$ above the ankle energy.
Therefore, the overall experimental energy spectrum of cosmic rays is described using the expression in
Eq.~(\ref{eqn:B2E1}),
For energies above
approximately
$5 \times 10^{19}$ [eV],
a drop in
the energy spectrum may appear due to the Greisen-Zatsepin-Kazumin (GZK) limit
\cite{Greisen,Zatsepin}.
\par
The theoretical ultraviolet cut-off in this paper is consistent with the following different experimentally observed phenomena:
\begin{itemize}
\setlength{\leftskip}{2ex}
\item[(1)] Consistency of the renormalization with the Einstein field equations\\
The Einstein field equations strongly require that all energies be included, and determines the space-time curvature.
In the renormalization process, the self-energy multiplied by a factor of $-1$ is added in the mass term of the Einstein field equations
as is given in Eq. (\ref{eqn:EeqR}) using Eq. (\ref{eqn:Smtm}).
Since this negative term cannot be regarded as a mass term, this term is moved to the other side of the equation and becomes the cosmological constant.
This is the origin of the cosmological constant in the theory.
\item[(2)] Moderate value of the cosmological constant\\
In the above renormalization,
Higgs self-energy, as a function of the cut-off energy, appears in the mass term of the Einstein field equations
in Eq. (\ref{eqn:EeqR}).
This procedure determines the relation between the cosmological constant and the cut-off energy.
The theory in this paper leads to our stable Universe.
If the cut-off energy is the Planck scale,
the cosmological constant leads to a rapid expansion of the universe.
Conversely, if the cut-off energy is the QCD scale, the cosmological constant
makes the size of the universe different from that of our Universe.
The cut-off seems not to be observed in the Large Hadron Collider (LHC).
\item[(3)] Inflation of the Universe\\
The cosmological constant in the theory presented in this paper is proportional to the energy density and decreases as the Universe expands. As derived in the preceding paper
\cite{Fuku18},
the time-dependence of the cosmological constant in this theory is proportional to $(t_{\rm U})^{-2}$, where $t_{\rm U}$ is the age of the Universe. Conversely, the inflation of the Universe then naturally occurs in the early Universe due to the large value of the cosmological constant, which is proportional to the energy density.
Since the cosmological constant depends on the matter density, the early Universe with the extremely high density of matter undergoes the inflation.
In Ref.
\cite{quint},
quintesence, which is not observed, is proposed to explain the Universe expansion. However, detailed properties of the quintesence, including the removal of ultraviolet divergences, are not presented.
\item[(4)] Non-detection of neutrinos above about the PeV-level\\
This theory presented in this paper suggests that elementary particles with light mass are not detected above the PeV-level, as mentioned in the Appendix of this paper; this includes neutrinos. Non-detection of neutrinos above the PeV-level is consistent with the IceCube experiments.
\end{itemize}
\par

\section{Conclusions}\label{sec:Sec4}

This study has investigated in detail the relation between the theoretical
ultraviolet cut-off energy and experimental data.
The theory presented in this paper is a fundamental cut-off theory, in which space-time continuum is divided into arbitrary-shaped space-time elements on the tangent space of the curved space-time continuum.
The wave function is expanded in terms of scalar plane waves 
using Lorentz/Poincar\'e covariant/invariant inner product of vectors.
In the normal scattering, the momentum lies within the cut-off energy.
Meanwhile, via the extremely weak interaction with the effective field, which is newly introduced,
scatterings rarely occur into the region between the cut-off energy and the Planck energy.
This formalism approaches the low-energy continuum theory, including gravity, in the limit of the small-length or high-energy cutoff.
Using current particle and cosmological data, a comparison
between the theoretical cut-off energy found in this work and
the experimentally observed spectrum of cosmic rays in the literature demonstrated the possibility that the cut-off energy corresponds to the experimental spectrum break,
called the knee at about 3 [PeV] in the case for the fixed cut-off energy.
The ultraviolet cut-off energy derived from the theoretically considered relation agrees with the experimentally observed tendencies, such as the stability of our Universe, inflation in the early Universe and non-detection of neutrinos above the energy of the order of PeV.
\par

\setcounter{equation}{0}
\renewcommand{\theequation}{\Alph{section}.\arabic{equation}}
\appendix
\def\thesection{Appendix:}
{\color{white}%
\section{
\hspace{-12.2ex}
{\color{black}
Appendix: Further Outlook
}
}
\color{black}

Cosmic rays contain neutrinos whose interactions with matter
in the Universe
are weak. The neutrino spectrum then indicates fundamental properties of elementary particles. In fact, a basic theoretical model is proposed in Ref.
\cite{WenYin},
aiming at the explanation of the IceCube experiments that show no neutrino detection above the PeV scale.
This model suggests
a new reaction absorbing a neutrino and its energy range by introducing elementary particles.
The energy range is from around the PeV-level to energies higher by some orders of magnitude,
and this energy range depends on the model
\cite{WenYin}.
By contrast, the theory in this paper leads to
that neutrinos are scattered via interactions with the effective fields, which are newly introduced, due to the long-range disorder in a system with the local periodicity of space-time elements obtained by dividing the space-time continuum.
The scattering
process is described using Feynman diagrams, which give the change
in a wave function with the aid of Green functions.
The process is represented in
Eq.~(\ref{eqn:dmat1})
below for a cosmic ray particle with
initial momentum $p^{\rm i}_{\rm c}$ and the associated wave function
$\psi_{p^{\rm i}_{\rm c}}$.
The momentum and associated wave function become $p^{\rm f}_{\rm c}$
and
$\psi_{p^{\rm f}_{\rm c}}$,
respectively, after the scattering.
At the first vertex, the wave function changes
due to the reaction with the effective disorder-induced field
of a polarization $\epsilon$.
Interaction strength denoted by the fine structure constant
is $\alpha_{\rm D}$,
and the cosmic ray particle absorb a boson with momentum
$k_{\rm D}=-G_{\mu \tilde{n}p}$
that is normal to $\epsilon$ in the spatial direction.
The interaction strength
$\alpha_{\rm D}$
is tiny when the deviation of the disordered alignment of space-time elements
from a regular periodic alignment
is small. At the second vertex, the cosmic ray particle
emits a boson with momentum
$k^{\prime}_{\rm D}=
-G_{\mu \tilde{n}^{\prime}q}$
of a polarization
$\epsilon^{\prime}$,
with the momentum
$k^{\prime}_{\rm D}$ being normal to $\epsilon^{\prime}$
in the spatial direction,
via the interaction with the disorder-induced field.
The Feynman diagram also contains a process in which the cosmic-ray particle
emit an effective disorder-induced boson with
momentum $k^{\prime}_{\rm D}$
before absorbing a boson with
momentum $k_{\rm D}$.
For the above process,
the transition matrix element, containing an adjoint quantity
$\bar{u}_{p^{\rm f}_{\rm c}}$,
is expressed by
\begin{eqnarray}
\nonumber
M^{\rm fi}
\propto
\alpha_{\rm D}
\bar{u}_{p^{\rm f}_{\rm c}}
\left[
(-i\slashed{\epsilon}^{\prime})
\frac{ i }
{
 \slashed{p}^{\rm i}_{\rm c}
+\slashed{k}_{\rm D}        -m_{\rm c}
}
(-i\slashed{\epsilon})
\right.
\\
+
\left.
(-i\slashed{\epsilon})
\frac{ i }
{
 \slashed{p}^{\rm i}_{\rm c}
-\slashed{k}_{\rm D}^{\prime}-m_{\rm c}
}
(-i\slashed{\epsilon}^{\prime})
\right]
u_{p^{\rm i}_{\rm c}}
\label{eqn:dmat1},
\end{eqnarray}
where
$p^{\rm i}_{\mu \rm c}$ and $m_{\rm c}$ are the initial momentum and the mass of the cosmic ray particle, respectively.
Conservation of momentum is satisfied by the disorder-induced
bosons in the above processes, as follows
$
 p^{\rm i}_{\rm c}+k         _{\rm D}
=p^{\rm f}_{\rm c}+k^{\prime}_{\rm D}.
$
\par
The scattering cross section of the cosmic-ray particle by the disorder-induced field
is derived from the corresponding cross section of the effective disorder-induced
field boson scattered by the cosmic-ray particle.
According to Ref.
\cite{BJDR}
concerning the scattering between a free particle and a massless boson, the total cross section
$\sigma_{\rm Lc}$
is approximated as
\begin{eqnarray}
\sigma_{\rm Lc}
=
\frac{\pi (\alpha_{\rm D})^2}
{k_{\rm D} m_{\rm c}}
\left[
\ln
\left(
\frac
{2k_{\rm D}}
{m_{\rm c}}
\right)
+\frac{1}{2}
\right]
\label{eqn:tcross},
\end{eqnarray}
where $k_{\rm D}$ and $m_{\rm c}$ in
Eq.~(\ref{eqn:dmat1})
and
the relation
$
 p^{\rm i}_{\rm c}+k         _{\rm D}
=p^{\rm f}_{\rm c}+k^{\prime}_{\rm D}
$
have been used.
The above cross section is roughly proportional to
$(m_{\rm c})^{-1}$.
For simplicity, the cut-off lengths along the
$x$, $y$, and $z$
directions in the real space-time are defined as
$\Delta_x$, $\Delta_y$,
and
$\Delta_z$,
respectively,
with $\Delta_x$$=\Delta_y$$=\Delta_z$.
When one cosmic ray particle travels along a trajectory of
length $L_{\rm c}$,
assuming that one scattering of the disorder-induced field occurs
within a volume $\Delta_x\Delta_yL_{\rm c}$,
the loss fraction of the cosmic ray particle is
\begin{eqnarray}
F_{\rm Lc}=
\left(
\frac{L_{\rm c}}{\Delta_z}
\right)
\frac{\sigma_{\rm Lc}}
{
\Delta_x\Delta_y
}
\label{eqn:CLfr1}.
\end{eqnarray}
Combining the above equation with
Eq.~(\ref{eqn:tcross})
yields
\begin{eqnarray}
\hspace{-1ex}
F_{\rm Lc}=
(\alpha_{\rm D})^2
\left(
\frac{L_{\rm c}}{\Delta_z}
\right)
\frac{1}
{\Delta_x\Delta_y}
\frac{\pi }
{k_{\rm D} m_{\rm c}}
\left[
\ln
\left(
\frac
{2k_{\rm D}}
{m_{\rm c}}
\right)
+\frac{1}{2}
\right]
\label{eqn:C2Lfr}.
\end{eqnarray}
The above loss fraction of the cosmic-ray particle is a function of the
coupling constant $\alpha_{\rm D}$, which corresponds to the fine-structure constant.
\par
Here, we consider the case for
$\Delta_x=\Delta_y=\Delta_z$,
corresponding to $3\times 10^{15}$ [eV], and
a trajectory length
$L_{\rm c}=1000$
[Mpc].
In Eq.~(\ref{eqn:tcross}),
we set
$\alpha_{\rm D}=
1.27 \times 10^{-28}$
and 
$k_{\rm D}=0.999 \times 10^{18}$
[eV], which is the energy of the disorder-induced boson
absorbed by the proton for energy $10^{18}$ [eV].
Then, the loss fraction $F_{\rm Lc}$ of the proton
in Eq.~(\ref{eqn:C2Lfr}) becomes
$10^{-2}$.
In this case, the loss fraction $F_{\rm Lc}$ is reduced to
$0.25 \times 10^{-4}$ for a trajectory length $L_{\rm c}=5$ [Mpc].
Since Eq.~(\ref{eqn:C2Lfr}) states that the cosmic-ray loss evaluated
using the scattering cross section is roughly proportional to the inverse of the
mass of the cosmic ray particle at an identical production rate, the loss of
a muon neutrino with a rest mass of less
than 0.17 [MeV] is larger by approximately four orders of magnitude than that of a proton.
Next, we treat the scattering of a cosmic ray photon by the
disorder-induced field (particle).
When the disorder-induced particle 
is a boson, the photon in the cosmic ray will create a fermion pair, which lose energy until the pair is annihilated and emit two photons with the energy of the fermion rest mass, in opposite directions.
When the disorder-induced
particle is an electron-like particle with mass smaller than an electron, the scattering cross section is larger by at least
approximately $1.8 \times 10^3$
than a proton, since the electron rest mass
is $1/1837$
of the proton mass.
\par
We note that in particular, elementary particles with negligible or zero rest
mass, such as neutrinos or photons, will be scattered with high scattering probabilities from the effective disorder-induced
field for all energies higher than the knee energy.
This will theoretically result in low detection possibility
of neutrinos and photons above approximately this value for the knee energy.
The obtained feature of the present theory is consistent with the experimental results that neutrinos and gamma rays are not observed
above approximately the knee energy \cite{Aartsen}, as mentioned in
Section
\ref{sec:Sec1}.
As a result, the composite fraction of heavy nuclei in cosmic rays becomes larger just above the knee energy.
These results are compatible with the experimentally observed composition of cosmic rays
\cite{Aartsen}.
In summary here,
the disorder-induced scattering of cosmic rays by the
long-range
disordered
alignment of the space-time elements,
which is obtained by dividing the space-time continuum,
theoretically leads to the knee.
This disorder-induced scattering gives rise to high scattering probabilities
for particles with tiny/zero rest mass.
Thus, the present theory is consistent with the absence of experimental detection
of neutrinos and photons for energies above approximately the knee energy.
\par
Here, the amount of disorder-induced bosons emitted in the Universe is evaluated, using the following experimental data.
The cosmic-ray flux at the Earth is defined using the
number of particles, measuring area (m$^2$) at the measuring center, solid angle in steradian (sr), time (year), and energy increment (GeV);
and the flux observed at $10^{18}$ [eV] is $\approx 3\times 10^{-14}$ [(m$^2$$\cdot$sr$\cdot$year$\cdot$GeV)$^{-1}$].
The Universe age, which is approximately 14 billion years old, is denoted as
$T_{\rm U} \approx 10^{10}$ [year].
Let us consider a cubic simulation box with a line length of
$L_{\rm U}=10^{10}$
[lightyear (ly)], through which cosmic rays pass during a time interval equal to the Universe age $T_{\rm U}$. As $L_{\rm U} \approx 10^{26}$ [m] with 1 [ly] $\approx 10^{16}$ [m], the total number of cosmic-ray particles that has reached the cube faces is given by
$\approx 6\times 3\times 10^{-14} \times (L_{\rm U})^2$
[(year$\cdot$GeV)$^{-1}$].
The production rate of the spatial number density of cosmic-ray particles in the cubic box is
$\approx 2\times 10^{-13} \times (L_{\rm U})^2/(L_{\rm U})^3$
$= 2\times 10^{-13} /L_{\rm U}=2\times 10^{-39}$
[(m$^3$$\cdot$year$\cdot$GeV)$^{-1}$].
The spatial number density of the cosmic-ray particles accumulated (produced)
in the cube at a time interval of $T_{\rm U}$, which is equal to the Universe age, is
$2 \times 10^{-29}$
[(m$^3$$\cdot$GeV)$^{-1}$].
For cosmic rays of the flux at $10^{18}$ [eV] in the energy region around $10^{18}$ [eV] with an energy increment of $10^{18}$ [eV], 
the spatial number density of the cosmic-ray particles accumulated
in the cube during a time interval $T_{\rm U}$ is roughly
$4 \times 10^{-20}$ [m$^{-3}$].
Meanwhile, it is assumed that half of the cosmic-ray particles are lost via scattering in a travel along the trajectory with the line length
$L_{\rm U}$
of the cubic simulation box.
The averaged energy emitted during disorder-induced scattering is set as 10$^8$ [GeV]. Then, the spatial density of the energy emitted during all scattering processes in the entire cube volume of $(L_{\rm U})^3$ during at a time interval $T_{\rm U}$ (the Universe age) amounts to
2 $\times 10^{-12}$ [GeV$\cdot$m$^{-3}$].
Meanwhile, the averaged spatial density of baryons in the Universe is
$\approx$ 0.3 [GeV$\cdot$m$^{-3}$].
\par
Furthermore, the disorder-induced boson, which is different from
bosons in the standard model,
interacts with all fermions. Then, the disorder-induced boson may belong to the matter similar to the dark matter, together with the disorder-induced fermion.
If the disorder-induced field with the symmetry, which is denoted as U$_{\rm D}$(1), is compatible with the standard model,
[SU(3)$]\times[$SU(2)$\times$U(1)] in the standard model may be replaced by 
[SU(3)]$\times$[SU(2)$\times$U(1)]$\times[$U$_{\rm D}$(1)], where
[\verb+ +] denotes that the group(s) separated by [\verb+ +] is (are) independent of the other group(s).

\begin{flushleft}

\end{flushleft}


\begin{thebibliography}{99}

\bibitem{Fuku84}
K.~Fukushima, {\it Phys.~Rev.~D} {\bf 30}, 1251 (1984). DOI:10.1103/PhysRevD.30.1251

\bibitem{Fuku14}
K.~Fukushima and H.~Sato, {\it Bulg.~J.~Phys.} {\bf 41{\rm{, no. 2}}}, 142 (2014). \\{U}RL: (article site)https://www.bjp-bg.com/paper1.php?id=696\\
(free pdf site)
https://www.bjp-bg.com/papers/bjp2014\_2\_142-171.pdf,
arXiv:1402.0450

\bibitem{Fuku16}
K.~Fukushima and H.~Sato, {\it Bulg.~J.~Phys.} {\bf 43{\rm{, no. 1}}}, 30 (2016). \\{U}RL: (article site) https://www.bjp-bg.com/paper1.php?id=799
\\(free pdf site) https://www.bjp-bg.com/papers/bjp2016\_1\_030-044.pdf,
arXiv:1501.04837

\bibitem{Fuku17}
K.~Fukushima and H.~Sato, {\it Int.~J.~Mod.~Phys.~A} {\bf 32}, 1730017 (2017). DOI:10.1142/S0217751X17300174\\
{U}RL: Free pdf is available at
\\https://www.worldscientific.com/doi/pdf/10.1142/S0217751X17300174

\bibitem{Fuku18}
K.~Fukushima and H.~Sato, {\it Eur.~Phys.~J.~C} {\bf 78}, 315 (2018). DOI:10.1140/epjc/s10052-018-5784-2

\bibitem{FemMW}
A.~R.~Mitchell and R.~Wait, {\it The Finite Element Method in Partial Differential Equations} (John Wiley \& Sons, New York, 1977).

\bibitem{BMS83}
C.~M.~Bender, K.~A.~Milton and D.~H.~Sharp, {\it Phys. Rev. Lett.} {\bf 51}, 1815 (1983). DOI:10.1103/PhysRevLett.51.1815

\bibitem{Weiss}
V.~S.~Weisskopf, {\it Det Kgl. Danske Videnskabernes Selskab. Mathematisk-fysiske Meddeleser} {\bf XIV{\rm, Nr. 6}}, 3 (1936).

\bibitem{Weiss39}
V.~S.~Weisskopf, {\it Phys. Rev.} {\bf 56}, 72 (1939). DOI:10.1103/PhysRev.56.72

\bibitem{Tomo46}
S.~Tomonaga, {\it Prog. Theor. Phys.} {\bf 1}, 27 (1946). DOI:10.1143/PTP.1.27

\bibitem{Tomo48}
S.~Tomonaga and J. R. Oppenheimer, {\it Phys. Rev.} {\bf 74}, 224 (1948). DOI:10.1103/PhysRev.74.224

\bibitem{Schwin}
J.~Schwinger, {\it Phys. Rev.} {\bf 76}, 790 (1949). DOI:10.1103/PhysRev.76.790

\bibitem{Feyn}
R.~P.~Feynman, {\it Phys. Rev.} {\bf 76}, 769 (1949). DOI:10.1103/PhysRev.76.769

\bibitem{Dys}
F.~J.~Dyson, {\it Phys. Rev.} {\bf 75}, 1736 (1949). DOI:10.1103/PhysRev.75.1736

\bibitem{Eins}
A.~Einstein, {\it {S}itzungsberichte der {K\"o}niglich {P}reu{\ss}ischen {A}kademie der {W}issenschaften ({B}erlin)} {\bf  VI}, 142 (1917).

\bibitem{Fried}
A.~Friedmann, {\it Z. Phys.} {\bf 10}, 377 (1922). DOI:10.1007/BF01332580

\bibitem{WenYin}
W.~Yin, {\it Chinese~Phys.~C} {\bf 43}, 045101 (2019).

\bibitem{Aartsen}
M. G. Aartsen {\it et al.} (IceCube), {\it Phys. Rev. D} {\bf 98}, 062003 (2018). DOI:10.1103/PhysRevD.98.062003

\bibitem{Yoshida}
S.~Yoshida and A.~Ishihara, {\it Phys. Rev. D} {\bf 85}, 063002 (2012). DOI:10.1103/PhysRevD.85.063002

\bibitem{BJDR}
J.~D.~Bjorken and S.~D.~Drell, {\it Relativistic Quantum Mechanics} (McGraw-Hill, New~York, 1964).

\bibitem{LanL}
L.~D.~Landau, E.~M.~Lifshitz, {\it The Classical Theory of Fields} {4th revised {E}nglish} edn. (Elsevier, Amsterdam, 1951).

\bibitem{higgsse}
URL: https://indico.cern.ch/event/691515/contributions/2970044/attachments/1688257/2715580/Terning-1.pdf

\bibitem{Birrell}
N.~D.~Birrell and P.~C.~W.~Davies, {\it Quantum Fields in Curved Space} (Cambridge~University~Press, Cambridge, 1982).

\bibitem{Tanabashi18}
M.~Tanabashi {\it et al.} (Particle~Data~Group), {\it Phys.~Rev.~D} {\bf 98}, 030001 (2018). URL: https://pdg.lbl.gov DOI:10.1103/PhysRevD.98.030001

\bibitem{Arason}
H.~Arason, D.~J.~Castano, B.~Kesthelyi, S.~Mikaelian, E.~J.~Piard, P.~Ramond and B.~D.~Wright, {\it Phys. Rev. D} {\bf 46}, 3945 (1992). DOI:10.1103/PhysRevD.46.3945

\bibitem{Kittel56}
C.~Kittel, {\it Introdoction to Solid State Physics} (Wiley, New York, 1956).

\bibitem{Kittel63}
C.~Kittel, {\it Quantum Theory of Solid} (John Wiley \& Sons, New York, 1963).

\bibitem{MEHandbook}
M. Abramowitz and A. Stegun (Ed.), {\it Handbook of Mathematical Functions with Formula, Graphs, and Mathematical Tables} (Dover Publications, New York, 1972).

\bibitem{Greisen}
K.~Greisen, {\it Phys. Rev. Lett.} {\bf 16}, 748 (1966). DOI:10.1103/PhysRevLett.16.748

\bibitem{Zatsepin}
G.~T.~Zatsepin and V.~A.~Kaz'min, {\it JETP~Lett.} {\bf 4}, 78 (1966).

\bibitem{quint}
P.~Ratra and L.~Peebles, {\it Phys.~Rev.~D} {\bf 37}, 3406 (1988). DOI:10.1103/PhysRevD.37.3406

\end{thebibliography}

\end{document}